\documentclass[12pt, reco]{article}
\usepackage{amssymb,amscd,amsmath,amsthm}
\usepackage{graphicx}
\newtheorem{theorem}{Theorem}[section]
\newtheorem{proposition}[theorem]{Proposition}

\newtheorem{definition}[theorem]{Definition}

\def\�{{\mathcal C}}

\def\ce{{\mathcal E}}

\def\Sl{{\mathcal S}}

\def\bn{{\mathbb N}}

\def\ot{\otimes}

\def\ffi{\varphi}

\def\tr{{\rm Tr}}
\def\Lb{\Lambda}

\def\El{\mathcal E}

\def\L{\Lambda}

\def\ffi{\varphi}

\def\Tr{\mathrm{Tr}}
\def\<{\langle}
\def\>{\rangle}
\def\1{\mathbf{1}}

\def\cal{\mathcal}

\def\L{\Lambda}

\def\id{{\bf 1}\!\!{\rm I}}
\begin{document}

 \begin{center}
{\Large {\bf Diagonalizability of quantum Markov States on trees}}\\[1cm]
\end{center}

\begin{center}
{\large {\sc Farrukh Mukhamedov}}\\[2mm]
\textit{ Department of Mathematical Sciences,\\
College of Science, United Arab Emirates University,\\
P.O. Box, 15551, Al Ain, Abu Dhabi, UAE}\\
E-mail: {\tt far75m@yandex.ru, \ farrukh.m@uaeu.ac.ae}
\end{center}

\begin{center}
{\large{\sc Abdessatar Souissi}}\\[2mm]
\textit{$^1$
College of Business Administration,\\
Qassim University, Buraydah, Saudi Arabia}\\
\textit{$^2$
  Preparatory Institute for Scientific and Technical Studies\\
Carthage University, Tunisia}\\
E-mail: {\tt a.souaissi@qu.edu.sa,  \ abdessattar.souissi@ipest.rnu.tn}
\end{center}

\begin{abstract}
We introduce quantum Markov states (QMS) in a general tree graph
$G= (V, E)$,  extending the  Cayley tree's case. We investigate
the   Markov property w.r.t. the finer structure of the considered
tree. The main result of this paper concerns the diagonalizability
of a locally faithful QMS $\varphi$ on a UHF-algebra $\mathcal
A_V$ over the considered tree by means of a suitable conditional
expectation into  a maximal abelian subalgebra. Namely, we prove
the existence of a Umegaki conditional expectation $\mathfrak E :
\mathcal A_V \to \mathcal D_V$ such that $$\varphi =
\varphi_{\lceil \mathcal D_V}\circ \mathfrak E.$$ Moreover, we
clarify the Markovian structure of the associated classical
measure on the spectrum of the diagonal algebra $\mathcal D_V$.

\end{abstract}

\section{Introduction }\label{intr}

The study of magnetic systems with competing interactions in
ordering is a fascinating problem of condensed matter physics
\cite{L86}. One of the most canonical examples of such systems is
frustrated Ising model which demonstrate a plethora of critical
properties \cite{CDS,MS}. Competing interactions (frustrations)
can result in new phases, change the Ising universality class, or
even destroy the order all together. Another interesting aspect of
the criticality in the frustrated Ising models is an appearance of
quantum critical points at spacial frustration points of model's
high degeneracy, and related quantum phase transitions \cite{CDS}.
The Ising models with frustrations can be thought as perturbation
of the classical Ising model. Recent studies show that to
investigate whole quantum system it is used "Matrix Product
States" and more generally to "Tensor Network States"
\cite{CV,Or}. This approach uses the density matrix
renormalization group (DMRG) algorithm which opened a new way of
performing the renormalization procedure in 1D systems and gave
extraordinary precise results. In this DMPG algorithm the
renormalization procedure takes explicitly into account the whole
system at each step. This is done by keeping the states of
subsystems which are relevant to describe the whole wave-function,
and not those that minimize the energy on that subsystem. This
kind of approach already has been considered in the literature as
Quantum Markov chains(QMC)\cite{[Ac74f]} which extends the
classical Markov chains into quantum level. Nowadays, the quantum
Markov chains have certain applications in solid-state physics,
quantum information theory and quantum statistical mechanics. The
reader is referred to
 \cite{AccKhreOhy,fannes,fannes2,Kum} for
recent development of the theory and its applications

On the other hand, in \cite{MBS161,MBS162} we have investigated a
quantum Markov chains associated to a particular class of the
Ising models with competing (commuting) interactions on the Cayley
trees. Recently, in \cite{[MuSou18]} we have established that the
above considered QMCs define a special class called
\textit{Quantum Markov States (QMS)}. Furthermore, description of
QMS has been carried out. It is worth to mention that introduced
QMS were considered over the Cayley trees, and investigated the
Markov property not only w.r.t. levels of the considered tree but
also w.r.t. to the interaction domain at each site, which is its
finer structure, and through  a family of suitable
quasi-conditional expectations so-called \textit{localized}
\cite{AccFi}. Furthermore, in \cite{MGS17,MGS19} we have
considered QMC corresponding to $XY$-models with competing Ising
interactions. It turns out that such kind of states do not
describe QMS. These are one of the first steps towards to
construction of a satisfactory theory of quantum analog of random
fields which is still one of the most interesting open problem in
quantum probability theory \footnote{First attempt towards quantum
Markov fields have been done in \cite{AccFid03}, \cite{Lib01},
\cite{AOM}.  Note that in the mentioned papers quantum Markov
fields were considered over multidimensional integer lattice
$\mathbb Z^d$. This lattice possesses the so-called amenability
property. Moreover, there do not exist analytical solutions on
such lattice. In \cite{[AccMuSou]} a  construction of quantum
Markov field was provided over arbitrary connected graph, but
concrete models are still missing. On the other hand, concrete
models based on quantum Markov chains on the Cayley tree were
studied in \cite{AccMuSa1,AccMuSa2,AccMuSa3}.}.

In the present paper, we are going to further study the structure
of QMS over arbitrary trees, and hence extend main result of
\cite{FF_FM,GZ} which concerns diagonalizability of
non-homogeneous QMS to general tree graphs. Namely, we show that
for every QMS $\varphi$ on the quasi-local algebra $\mathcal A$
there exists a suitable maximal abelian subalgebra $\mathcal D$
and a classical Markov measure $\mu$ on the spectrum
$\mathrm{spec}(\mathcal D)$ and a suitable conditional expectation
$\mathfrak E: \mathcal A \to \mathcal D$ such that $\varphi =
\varphi_\mu\circ \mathfrak E$, the state $\varphi_\mu$ being the
restriction of $\varphi$ on the diagonal algebra $\mathcal D$.
Moreover, the Markov property was expressed w.r.t. the finer
structure of the considered tree. This result will allow us to
find entropy of QMS in general setting, and may further open new
insight in the theory of quantum Markov fields. Moreover, the
result of the paper could be applied to the investigation of
quantum systems governed by QMS over complex networks
\cite{D2010}.

Let us briefly outline an organization of the this paper. Section
2 is devoted to preliminary notions and fact about trees and
algebras of observables. Furthermore, in section 3, we give the
definitions of quantum Markov chains and states over trees through
an appropriate Markov property. Then in section 4, we investigate
the Markov property associated to a given quantum Markov state on
the infinite tensor product of full matrix algebras. Namely, we
determine a factorisation of the associated potentials, based on
the finer structure of the considered tree. In section 5, we prove
the diagonalizability theorem of quantum Markov states, and in the
final section 6, the markovianity of the associated measure $\mu$
on the spectrum is established.

\section{Preliminaries}

\subsection{Trees}

Let us consider an infinite connected graph $G=(V, E)$, here $V$
stands for the set of vertices and $E$ is the set of edges. Two
vertices $x$ and $y$ are called {\it nearest neighbors} and they
are denoted by $l=<x,y>$ if there exists an edge connecting them.
In what follows, any connected graph $G$ is not containing no
cycles, is called \textit{tree}. Roughly speaking, from every
vertex $x\in V$ issues a finite number of edges. If this number is
constant, say equal to $k\geq 1$, for every vertex, then the tree
is called \textit{Cayley tree} of order $k$. Any collection of
pairs $<x,x_1>,\dots,<x_{d-1},y>$ in $G$ is called a {\it path}
from the point $x$ to the point $y$. The distance $d(x,y), x,y\in
V$, on the tree, is the length of the shortest path from $x$ to
$y$.

Fix a root $x_0\in  V$, and denote
$$\Lambda_n= \{x\in V \;  : \;  d(x,
x_0) =n\}, \ \ \Lambda_{[0,n]} = \bigcup_{k=0}^{n}\Lambda_k.$$
Notice that the set $\Lambda_n$ is finite. Due to tree structure,
each element $x\in \Lambda_n$ can be joined to $x_0$ through a
unique path from $x_0$ to $x$.

To each vertex  $x\in \Lambda_n$, we associate the set of its {\it direct successors} as follows
\begin{equation}\label{S(x)def}
\overrightarrow{S}(x) = \{y\in \Lambda_{n}^c \; \mid \;
d(x,y)=1\}, \quad  \quad k_x= |\overrightarrow{S}(x)|
\end{equation}

 The set $\overrightarrow{S}(x)$ is finite, and  it is eventually empty for some vertices of $\Lambda_n$ but not for all of them.
  Moreover, the sets $\overrightarrow{S}(x), \, x\in \Lambda_n$ form a partition of  $\Lambda_{n+1}$ i.e.
\begin{equation}\label{Ln_Vn+1}
\Lambda_{n+1} = \bigsqcup_{x\in \Lambda_n}\overrightarrow{S}(x)
\end{equation}
 where  $\bigsqcup$ denotes a disjoint union.

We recall that if one reduces the study to the homogenous case
i.e. $k_n =k$ for a unique integer $k$,
 the graph $G$ reduces to the semi-infinite Cayley tree $\Gamma^k_+$ of order $k$  studied in \cite{MBS161,MBS162}.

In what follow,  a coordinate structure will be set up w.r.t.
levels $(\Lambda_n)_n$ as follows:

$$\Lambda_0 = \{x_0\} \quad ; \quad x_0  : =(0) $$

Having defined a coordinate structure on $\Lambda_n$, let us
denote
$$
\overrightarrow\Lambda_n =\left\{ x_{\Lambda_n} (1),
x_{\Lambda_n}(2), \cdots,x_{\Lambda_n}(|\Lambda_n|) \right\}
$$

Now for each $\in \overrightarrow\Lambda_n$, we consider a
(random) enumeration on its set of direct successors as follows
$$
\overrightarrow S(x) = \left\{ \bigl(x ,\;  1\bigr),\bigl(x, \;
2\bigr), \cdots , \bigl( x,\; k_{x}\bigr) \right\}
$$
Taking into account \eqref{Ln_Vn+1}, one gets coordinate structure on the level $\Lambda_{n+1}$ based on the enumeration on  $\Lambda_n$.
In this way,   an enumeration on the full vertex set $L$ is defined.

%

\subsection{Inclusions of $C^*$-algebras}

In this section, we recall  some well-known facts about inclusions
of finite dimensional $C^*$-algebras.

 In what follows  $ \mathcal  A$
stands for a finite dimensional $C^*$-algebra, and assume $
\mathcal B$ is a subalgebra of $ \mathcal A$. A projection $p$ in
$\mathcal A$ is said to be minimal if for each projection $q\in
\mathcal A$
$$
q< p \Rightarrow q= 0
$$

Consider finite sets $\{p_i\}_i$ of all minimal central
projections, respectively of $\mathcal A$ and $\mathcal B$
 such that
\begin{equation}\label{centralproj}
\sum_i p_i = \id \quad ; \quad \sum_j q_j = \id
\end{equation}

From \eqref{centralproj}  one gets
 \begin{equation}\label{U=Uij}
 \mathcal A = \sum_{i}p_i\mathcal A \quad ; \quad \mathcal B = \sum_{j}q_j\mathcal B
 \end{equation}

Notice that $p_i\mathcal Ap_i = p_i \mathcal A$ and $q_j\mathcal B
q_j = q_j \mathcal B$.  Therefore, we denote  $\mathcal A_i :=
p_i\mathcal A \, ; \, \mathcal B_j := q_j\mathcal B \, ; \,
\mathcal A_{ij} = q_jp_i\mathcal  M p_i q_j \, ; \,  \mathcal
B_{ij} = q_jp_i \mathcal B p_i q_j$. For each $i,j$ one has
inclusion $\mathcal B_{ij} \subseteq \mathcal A_{ij}$ of finite
factors. Then
 \begin{equation}\label{tensor_decompij}
 \mathcal A_{ij}  \sim \mathcal B_{ij} \otimes\bar{\mathcal B}_{ij}
 \end{equation}
  for an other finite dimensional factor $\bar{\mathcal B}_{ij}$.

Let us consider the canonical traces $\Tr_{\mathcal A}, \quad
\Tr_{\mathcal B}$, on ${\mathcal A}$ and ${\mathcal B}$,
respectively, which take unit values on minimal projections.
Taking into account the identifications \eqref{U=Uij} and
\eqref{tensor_decompij}, one finds
$$
\Tr_{\mathcal A} = \bigoplus_{i,j}\Tr_{\mathcal B_{ij}}\otimes \Tr_{\bar{\mathcal B}_{ij}}
$$

From the above considerations, the equality
\begin{equation}
E_{\mathcal B}^{\mathcal A} : = \bigoplus_{i,j} (\mathrm{id}_{\mathcal B_{i,j}}
\otimes \Tr_{\bar{\mathcal B}_{ij}})
\end{equation}
defines a linear completely positive and $(\Tr_{\mathcal U},
\Tr_{\mathcal B})$-preserving
 map from $\mathcal A$ onto $\mathcal B$.

 Let  $\varphi$ be given a state on the algebra $\mathcal A$, together with
 its restriction $\varphi \lceil \mathcal B$. Consider the corresponding Radon-Nikodym
  derivatives $T^{\varphi}_{\mathcal A}, T^{\varphi}_{\mathcal B}$ w.r.t.
    the canonical traces $\Tr_{\mathcal A}, \Tr_{\mathcal B}$ respectively. Then
 \begin{equation}
 T_{\mathcal B}^{\varphi} = E_{\mathcal A}^{\mathcal B}( T_{\mathcal A}^{\varphi})
 \end{equation}

Recall that a Umegaki \textit{conditional expectation}
$E:\mathcal{A}\to \mathcal{B}$ is a norm-one projection of the
$C^*$-algebra $\mathcal{A}$ onto a $C^*$-subalgebra $\mathcal{B}$
(with the same identity $\id$). The map $E$ is automatically a
completely positive, identity-preserving $\mathcal{B}$-module map
\cite{Str}. If $\mathcal{A}$ is a matrix algebra, then the
structure of a conditional expectation is well-known \cite{ACe}.

Let is recall some facts. Assume that $\mathcal{A}$ is a full
matrix algebra, and consider the (finite) set of $\{P_i\}$ of
minimal central projections of the range $\mathcal{B}$  of $E$, we
have
$$
E(x)=\sum_iE(P_ixP_i)P_i.
$$
Then $E$ is uniquely determined by its values on the reduced
algebras
$$
\mathcal{A}_{P_i}:=P_i\mathcal{A}P_i=N_i\otimes\bar N_i,
$$
where $N_i\sim \mathcal{B}_{P_i}:=\mathcal{B}P_i$ and $\bar
N_i:=\mathcal{B}'P_i$ (here the commutant $\mathcal{B}'$ is
considered relative to $\mathcal{A}$). Moreover, there exist
states $\phi_i$ on $\bar N_i$ such that
\begin{equation}\label{Exp1}
E(P_i(a\otimes \bar a)P_i)=\phi_i(\bar a)P_i(a\otimes \id)P_i.
\end{equation}
For the general theory of operator algebras we refer to \cite{BR,
Str}.

\section{Quantum Markov chains and states on trees }

Let $G=(V,E)$ be an infinite tree.  To each vertex $x\in V$, one
associates a finite $C^*$-algebra $\mathcal A_x$.
 For $\Lambda \subseteq V$ we set the local algebra $\mathcal A_\Lambda = \bigotimes_{x\in \Lambda}\mathcal A_x$.
Notice that for $\Lambda_1\subset \Lambda_2$ one has
$$\mathcal A_{\Lambda_1} \cong \mathcal A_{\Lambda_1}\otimes \id_{\Lambda_2\setminus\Lambda_1 }\subset \mathcal A_{\Lambda_2}$$
 $$\mathcal A_\Lambda = \bigotimes_{x\in \Lambda}\mathcal A_x.$$
By $\mathcal A_V$ we denote the inductive limit of $C^*$-algebras,
that is,
$$
\mathcal A_V = \lim_{\Lambda \uparrow V}\mathcal A_\Lambda
$$
 Since the C$^*$--algebra $\mathcal A_V$ is isomorphic to the algebra $\bigotimes_{x\in L}\mathcal A_{x}$,
 the algebra $\mathcal A_x$ can be viewed as subalgebra of the algebra $\mathcal A_V$  trough the following embedding
 \begin{equation}
 j_x : a \in \mathcal A_x \mapsto j_x(a) = a\otimes \id_{\{x\}^c}
 \end{equation}
More generally, for every $\Lambda\subset_{fin} L$ we define
$$
 j_{\Lambda} = \bigotimes_{x\in \Lambda}j_x
 $$
 To simplify the notations, in the following we will often identify each $\mathcal A_\Lambda$
 to the subalgebra $j_\Lambda(\mathcal A_\Lambda)$ of $\mathcal A_V$, through the identification
 $$
 \mathcal A_\Lambda \equiv \mathcal A_\Lambda\otimes\id_{\Lambda^c}
 $$
 In this notations we set the following local subalgebra defined by
\begin{equation}
\mathcal A_{L, loc} = \bigcup_{\Lambda\subset L}\mathcal
A_{\Lambda}
\end{equation}
which generates the algebra $\mathcal A_{L}$.

Let $\{\mathcal B_{\Lambda }\}_{\Lambda\subseteq L}$  be given a
net of local algebras such that
\begin{equation}
\mathcal A_{\Lambda_{[0,n]}} \subset \mathcal
B_{\Lambda_{[0,n+1]}} \subset \mathcal A_{\Lambda_{[0,n+1]}}
\end{equation}

Let  $\mathcal D_{\Lambda}$ be a maximal abelian subalgebra of
$\mathcal B_{\Lambda}$. Then the C$^*$--inductive limit
\begin{equation}
\mathcal D = \overline{(\lim_{\Lambda\uparrow L} \mathcal
D_{\Lambda})}
\end{equation}
is an abelian C$^*$--subalgebra of $\mathcal A_V$ and it is called
\textit{diagonal algebra}. The reader is referred to \cite{BR} for
a detailed study of the subject.

Let us consider a triplet ${\cal C} \subset {\cal B} \subset {\cal
A}$ of unital $C^*$-algebras. Recall \cite{ACe} that a {\it
quasi-conditional expectation} with respect to the given triplet
is a completely positive (CP), identity-preserving linear map $\ce
\,:\, {\cal A} \to {\cal B}$ such that
$$ \ce(ca) = c \ce(a),  \quad \forall a\in {\cal A},\, c \in {\cal C}.
$$

\begin{definition}\cite{[AcFiMu07],AOM}\label{QMCdef}
Let $\varphi$ be a state on $\mathcal{A}_V$. Then $\ffi$ is called
a {\it  quantum Markov chain}, associated to $\{\L_n\}$, if for
each $n\in\bn$ there exist a quasi-conditional expectation
$\ce_{\Lambda_{n]}}$ with respect to the triple
$\mathcal{A}_{{\Lambda}_{n-1]}}\subseteq
\mathcal{A}_{\Lambda_n}\subseteq\mathcal{A}_{\Lambda_{n+1]}}$ and
an initial state $\rho_0\in S(\mathcal{A}_{\L_0})$ such that
\begin{equation*}
\varphi = \lim_{n\to\infty} \rho_0\circ \ce_{\Lambda_{0]}}\circ
\ce_{\Lambda_{1]}} \circ \cdots \circ \ce_{\Lambda_{n]}}
\end{equation*}
in the weak-* topology.
\end{definition}

 \begin{definition}\label{QMS1}\cite{[AcFiMu07]}
A quantum Markov chain $\ffi $ is said to be quantum Markov state
with respect to the sequence $\{\ce_{\Lambda_{j]}}\}$ of
quasi-conditional expectations if one has
\begin{equation}\label{Markov_state_eq}
\ffi_{\lceil \mathcal{A}_{\Lambda_{j]}}}\circ \ce_{\Lambda_{j]}}=\ffi_{\lceil \mathcal{A}_{\Lambda_{j+1]}}}, \ \ j\in\bn
\end{equation}
\end{definition}

One can check that the above Markov property (\ref{Markov_state_eq}) can be formulated using
a sequence of global quasi--conditional expectations, or equally
well by sequences of local or global conditional expectations. By
taking $e_{n}:=\ce_{\Lambda_{n]}}\lceil_{\mathcal{A}_{\Lambda_{[n,n+1]}}}$, it
will be enough to consider the ergodic averages
$$
\ce^{(n)}:=\lim_m\frac{1}{m}\sum^{m-1}_{h=0}(e_{n})^{h}\,,
$$
which give rise to a sequence of two--step conditional expectations,
called {\it transition expectations} in the sequel.

 For $j>0$, we define the conditional expectation $E_{j}$ from $\mathcal{A}_{\Lb_{j+1}}$
  into $\mathcal{A}_{\Lb_{j}}$  by:
 $$
 E_{j}\left(a_{\Lambda_0}\ot\cdots \ot  a_{\Lambda_j}\ot a_{\Lambda_{j+1}}\right)
 =a_{\Lambda_0}\ot\cdots \ot a_{\Lambda_{j-1}}\ot \El^{(j)}\left(a_{\Lambda_{j}}\ot a_{\Lambda_{j+1}}\right)
 $$

Using the argument of \cite{AccFi} one can prove the following
result.

\begin{proposition}\label{qmc_eqUme_Qua}
Let $\varphi$ be a state on the $\mathcal{A}_V$. The following
statements are equivalent:
\begin{itemize}
\item[(i)] $\varphi$ is a quantum Markov state;
\item[(ii)] the properties listed in Definitions \ref{QMCdef} and \ref{QMS1} are
satisfied if one replaces the quasi--conditional expectations
$\ce_{\Lambda_n}$ with Umegaki conditional expectations $E_{n}$.
\end{itemize}
\end{proposition}

The next result describes the quantum Markov states.

\begin{theorem}\label{caracofmarkovstate}
Let $\ffi\in \Sl(\mathcal{A}_V)$. Then $\ffi$ is a quantum Markov state
w.r.t the sequence of transition expectations
$\{\El^{(j)}\}_{j\geq0}$ if and only if
   \begin{equation}\label{Mp1}
   \ffi(a)=\ffi\left(\El^{(0)}\left(a_{\Lambda_0}\ot\cdots \ot \El^{(n-1)}
   \big(a_{\Lambda_{n-1}}\ot\El^{(n)}\left(a_{\Lambda_n}\ot a_{\Lambda_{n+1}}\right)\cdots\right)\right)
   \end{equation}
  for every $n\in \bn$, and $a=a_{\Lambda_0}\ot \cdots a_{\Lambda_n}\ot a_{\Lambda_{n+1]}}$ any linear generator of $\mathcal{A}_{\Lb_{n+1}}$,
  with $a_{\Lambda_j}\in \mathcal{A}_{\Lambda_j}$ for $j=1\cdots n+1.$
\end{theorem}

The proof uses the same argument used in \cite{[MuSou18]} where a
similar result has been established in the case of the Cayley
trees.

\section{Factorization of potentials associated to QMS on trees }

In sequel, we take $\mathcal{A}_x = M_{d_x}(\mathbb C$) where   $d_x\in\mathbb N$ for all $x\in L$. Let $\varphi$ be a Markov state on $\mathcal{A}$ together with its sequence $\mathcal E_{\Lambda_{n]}}$ of quasi-conditional
 expectations  w.r.t. the triplet  $\mathcal{A}_{\Lambda_{n-1]}}\subset  \mathcal{A}_{\Lambda_{n]}}\subset \mathcal{A}_{\Lambda_{n-1]}}$.

Taking into account Proposition \ref{qmc_eqUme_Qua},  without loss
of generality, one assumes  that  $\mathcal E_{\Lambda_{n]}}$ is a
(Umegaki) conditional expectation rather than quasi-conditional
expectation.

Define
$$
\mathcal E_{\Lambda_{[n,n+1]}}:={\mathcal E_{\Lambda_{n]}}}_{\Big\lceil\mathcal{A}_{\Lambda_{[n,n+1]}}}:=\hbox{the restriction of $\mathcal E_{\Lambda_{n]}}$ on }
\mathcal{A}_{\Lambda_{[n,n+1]}}
$$

From  \eqref{Markov_state_eq}, one gets
\begin{equation}\label{E[n,n+1]restr}
\varphi_{\big\lceil \mathcal{A}_{\Lambda_n}}\circ \mathcal E_{\Lambda_{n}} = \varphi_{\big\lceil \mathcal{A}_{\Lambda_n}}\circ \mathcal E_{\Lambda_{[n,n+1]}}
\end{equation}

In the sequel, we always assume that the map $\mathcal
E_{\Lambda_{[n,n+1]}}$ is assumed to satisfy the following
\textit{localization property}
  \begin{equation}\label{localizationCE}
  \mathcal E_{\Lambda_{[n,n+1]}}\bigl(\mathcal{A}_{\{x\}\cup \overrightarrow S(x)} \bigr) \subseteq \mathcal{A}_x
  \end{equation}

We notice that this property \eqref{localizationCE} plays a key
role for the integral decomposition of QMS since takes into
account finer structure of conditional expectations and
filtration.

Denote
$$
\mathcal E^{x}:= {\mathcal E_{\Lambda_{[n,n+1]}}}_{\big\lceil \mathcal{A}_{\{x\}\cup \overrightarrow S(x)}}
$$

 In the special case when $\overrightarrow S(x) =\varnothing$, one can choose  $\mathcal{A}_{\overrightarrow S(x)} \equiv \mathbb C\id $ and $\mathcal E^x \equiv {\rm id}_{\mathcal{A}_x}$.

  From  \eqref{localizationCE}, the map $\mathcal E^x$ defines a Umegaki conditional expectation from $\mathcal{A}_{\{x\}\cup \overrightarrow S(x)}$ into $\mathcal{A}_x$. Moreover, one gets
  $$
  \mathcal E_{\Lambda_{[n,n+1]}} = \bigotimes_{k=1}^{|\Lambda_n|} \mathcal E^{x_{\Lambda_n}(k)}
  $$
and then the range $\mathcal R(\mathcal E_{\Lambda_{[n,n+1]}})$  of $ \mathcal E_{\Lambda_{[n,n+1]}}$ has the following decomposition
\begin{equation}\label{Range_E_n_decomp}
\mathcal R(\mathcal E_{\Lambda_{[n,n+1]}}) = \bigotimes_{k=1}^{|\Lambda_n|} \mathcal R(\mathcal E^{x_{\Lambda_n}(k)} )
\end{equation}

In the following, we are going to find an equivalent formulation
of the compatibility condition \eqref{Markov_state_eq} for quantum
Markov state through the local conditional expectations
$\{{\mathcal E}^x : \; x\in L\}$.

Let   $\{P^x_{\omega_x}\}_{\omega_x\in \Omega_x}$ be the set of all minimal  central projections of  the range $\mathcal R_x := \mathcal R(\mathcal E_x)$ of the conditional expectation $\mathcal E^x$ . Recall that
$$
\sum_{\omega_x\in\Omega_x}P_{\omega_x}^x = 1
$$
Put
$$
\mathcal N_x : = \bigoplus_{\omega_x\in\Omega_x} P_{\omega_x}^x\mathcal{A}_x P_{\omega_x}^x
$$

For each $x\in V$, one defines
 \begin{equation}\label{E_projection }
 E^x(a) := \sum_{\omega_x\in\Omega_x} P_{\omega_x}^x a P_{\omega_x}^x
\end{equation}
and
\begin{equation}\label{E_Vambad_n}
 E_{[n, n+1]} := \bigotimes_{x\in \Lambda_n} E^x \quad ; \quad  E_{\Lambda_{n]}} = \bigotimes_{k=0}^n  E_{[k, k+1]}
\end{equation}

\begin{proposition} \label{prop_phiE}
For the same notations as above, the following assertions holds
true:
\begin{enumerate}
\item[(i)] $\mathcal E_{\Lambda_{n]}} \circ E_{\Lambda_{n]}}
=E_{\Lambda_{n]}}\circ \mathcal E_{\Lambda_{n]}}  = \mathcal
E_{\Lambda_{n]}}$;
 \item[(ii)] $\varphi = \varphi_{\big\lceil
\Lambda_{n]}}\circ E_{\Lambda_{n]}}$.
\end{enumerate}
\end{proposition}
The reduced algebra $P_{\omega_x}\mathcal{A}_x P_{\omega_x}$ can be written as tensor product of two finite factors $N_{\omega_x}$ and $\bar N_{\omega_x}$ as follows
\begin{equation}
P_{\omega_x}\mathcal{A}_x P_{\omega_x} = N_{\omega_x}\otimes\bar N_{\omega_x}
\end{equation}
Let $\Lambda\subset V$  be a finite subset, the spectrum of the algebra $\mathcal R_\Lambda : = \bigotimes_{x\in\Lambda}\mathcal R(\mathcal E^x)$ is given by
$$\Omega_{\Lambda} = \prod_{x\in\Lambda}\Omega_x$$
Then its elements are of the form $\Lambda_\Lambda = (\omega_x)_{x\in \Lambda}$ and their associated minimal central projections of $\mathcal R_\Lambda$ are of the form
$$P_{\omega_\Lambda}: = \bigotimes_{x\in\Lambda}P_{\omega_x}$$

$$
\mathcal N_{\Lambda} = \bigoplus_{\omega_\Lambda\in \Omega_\Lambda}\bigotimes_{x\in\Lambda} N_{\omega_x}\otimes\bar N_{\omega_x}
$$
In particular, the algebra $\mathcal N_{\Lambda_{n]}}$ can be written as follows
\begin{equation}\label{algebra_U}
\mathcal N_{\Lambda_{n]}} = \bigoplus_{\omega\in \Omega_{n]}}\left( N_{\omega_{x_0}}\bigotimes_{k=0}^{n-1}\bigotimes_{x\in \Lambda_k} \bigl(\bar N_{\omega_{x}}\otimes N_{\omega_{(x,1)}}\otimes N_{\omega_{(x,2)}}\otimes \cdots \otimes N_{\omega_{(x,k_x)}}\bigr)\bigotimes_{y\in \Lambda_n}\bar N_{\omega_{y}}\right)
\end{equation}

Let us denote
\begin{equation}\label{N_V}
\mathcal N_V : = \overline{\bigcup_{n\in\mathbb N} \mathcal N_{\Lambda_{n]}}}^{C^\ast}
\end{equation}
Consider the family of potentials  $\bigl\{ h_{\mathcal A_{\Lambda}}\bigr\}_{\Lambda\subset_{fin}V}$  associated to the state $\varphi_{\Lambda}:= \varphi_{\big\lceil{\mathcal{A}_{\Lambda}}}$ through the formula:

\begin{equation}
\varphi_\Lambda(\cdot) = \Tr(e^{-h_{\mathcal{A}_{\Lambda}}}\cdot)
\end{equation}
If $\Lambda = \{x\}\cup \overrightarrow S(x)$, with $x\in V$, then
the potential $h_{\{x\}\cup \overrightarrow S(x)}$ has the
following form
\begin{equation}
h_{\mathcal{A}_{\{x\}\cup \overrightarrow S(x)}} =\bigoplus_{\omega \in \Omega_{\{x\}\cup \overrightarrow S(x)}}\left( h^x_{\omega_x}\otimes\bigl( h^{x}_{(\omega_{x} , \omega_{(x,1)}, \omega_{(x,2)},  \cdots , \omega_{(x,k_x)})}\bigr)\otimes \bigl(\hat h^{(x,1)}_{\omega_{(x,1)}}\otimes \hat h^{(x,2)}_{\omega_{(x,2)}}\cdots \otimes \hat h^{(x,k_x)}_{\omega_{(x,k_x)}}\bigr)\right)
\end{equation}
where  self-adjoint operators $h^x_{\omega_x},\;
h^{x}_{(\omega_{x} , \omega_{(x,1)}, \omega_{(x,2)},  \cdots ,
\omega_{(x,k_x)})}$ and $ \hat h^{(x,j)}_{\omega_{(x,j)}}$ are
localized
 in the factors $N_{\omega_x}^x,\;  \bar N_{\omega_x}^x\otimes N^{(x,1)}_{\omega_{(x,1)}}\otimes N^{(x,2)}_{\omega_{(x,2)}}\otimes\cdots\otimes N^{(x,k_x)}_{\omega_{(x,k_x)}}$ and $\bar N^{(x,j)}_{\omega_{(x,j)}}$
respectively.

Therefore, the potential $h_{\Lambda_{n]}}$ has the following decomposition
\begin{equation}
h_{\mathcal{A}_{n]}} = \bigoplus_{\omega\in \Omega_{n]}}h^{x_0}_{\omega_{x_0}}\bigotimes_{k=0}^{n-1}\bigotimes_{x\in \Lambda_k}\bigl( h^{x}_{(\omega_{x} , \omega_{(x,1)}, \omega_{(x,2)},  \cdots , \omega_{(x,k_x)})}\bigr)\bigotimes_{y\in \Lambda_n}\hat h_{\omega_{y}}^{y}
\end{equation}
where the operators $h^{x_0}_{\omega_{x_0}},\; h^{x}_{(\omega_{x}
, \omega_{(x,1)}, \omega_{(x,2)},  \cdots , \omega_{(x,k_x)})}$
and $  \hat h^{y}_{\omega_{y}}$ are localized in the factors
 $N_{\omega_{x_0}}^{x_0};\quad \bar N_{\omega_x}^x\otimes N^{(x,1)}_{\omega_{(x,1)}}\otimes N^{(x,2)}_{\omega_{(x,2)}}\otimes\cdots\otimes N^{(x,k_x)}_{\omega_{(x,k_x)}}$ and $   \bar N^{y}_{\omega_{y}}$ respectively.\\

 Put
 $$
 H_{x} = \sum_{\omega_{x}\in\Omega_{x}}P^{x}_{\omega_{x}}(h^{x}_{\omega_{x}}\otimes\id )P^{x}_{\omega_{x}},
  \quad \hat H_{x}:
  = \sum_{\omega_{x}\in\Omega_{x}}P^{x}_{\omega_{x}}(\id \otimes \hat h^{x}_{\omega_{x}} )P^{x}_{\omega_{x}}
  $$

$$
H_{\{x\}\cup \overrightarrow S(x)} : =\sum_{\omega \in \Omega_{\{x\}\cup \overrightarrow S(x)}}P^{x}_{\omega}\bigl( \id \otimes  h^{x}_{(\omega_{x} , \omega_{(x,1)}, \omega_{(x,2)},  \cdots , \omega_{(x,k_x)})}\otimes \id \bigr)P^{x}_{\omega}
$$
 where
$$
 P^{x}_{\omega}:=  P^x_{\omega_x}\otimes P^{(x,1)}_{\omega_{(x,1)}}\cdots \otimes P^{(x,k_x)}_{\omega_{(x,k_x)}}.
$$
 The self-adjoint operators $H_{x}$ localized in $\mathcal A_x$ and $H_{\{x\}\cup \overrightarrow S(x)}$ localized in $\mathcal A_{\{x\}\cup \overrightarrow S(x)}$ satisfy the following commutation relations
 \begin{equation}\label{comm1}
 [H_x ,  H_{\{x\}\cup \overrightarrow S(x)} ] = [H_{\{x\}\cup \overrightarrow S(x)} , \hat H_{(x,i)} ] =0 \quad ; \quad x\in V , i\in \{1, \dots, k_x\}
 \end{equation}
 \begin{equation}\label{comm2}
  [H_x , \hat H_y] = [H_{\{x\}\cup \overrightarrow S(x)} , H_{\{y\}\cup S(y) }]=0 \quad ; \quad x,y\in V
 \end{equation}
and
\begin{equation}\label{h_mdecomposition}
h_{\mathcal{A}_{n]}} = H_{x_0}  + \sum_{k=0}^{n-1}\sum_{x\in \Lambda_k} H_{\{x\}\cup \overrightarrow S(x)}  + \sum_{y\in\Lambda_n }\hat H_{y}
\end{equation}
for each $n\in\mathbb N$.

\section{Diagonalizability of Markov states on trees } \label{diag_section}

In this section, we prove the main result of this paper which is
given in the following theorem.

\begin{theorem}\label{main_thm}
Let $\varphi\in \mathcal S(\mathcal{A}_V)$ be a quantum Markov
state. Then there exists a diagonal algebra $\mathcal
D_V\subset\mathcal N_V$, a classical Markov measure $\mu$ on
$\mathrm{spec}(\mathcal D_V)$ and a Umegaki conditional
expectation $\mathfrak E: \mathcal{A_V} \to \mathcal D_V$ such
that $\varphi = \varphi_\mu\circ \mathfrak E$, where $\varphi_\mu$
is the state on $\mathcal D_V$ corresponding to $\mu$.
\end{theorem}

\proof Given  $x\in \Lambda_n$,  by $\mathcal R_x$ we denote the
range of the conditional expectation $\mathcal E_x$. Let
$$\mathcal R_n = \bigotimes_{x\in \Lambda_n}\mathcal R_x.$$ Consider
the algebra $\mathcal N_{\Lambda_{n]}}$ defined by
\eqref{algebra_U}. Given $\omega_{\{x\}\cup \overrightarrow
S(x)}\in \Omega_{\{x\}\cup \overrightarrow S(x)}$, by
$D_{\omega_{\{x\}\cup \overrightarrow S(x)}}$ we denote a maximal
abelian sub-C$^{\ast}$-algebra  of $\bar N_{\{\omega_x\}}\otimes
N_{\omega_{(x,1)}}\cdots \otimes N_{\omega_{(x,k_x)}}$ containing
$ h^{x}_{(\omega_{x} , \omega_{(x,1)}, \omega_{(x,2)},  \cdots ,
\omega_{(x,k_x)})}$. Similarly, $D_{\omega_{x_0}}$ denotes  a
maximal abelian sub-C$^{\ast}$-algebra of $N_{\omega_{x_0}}$.

Let us define
 \begin{equation}\label{decomp_D_Vn}
\mathcal D_{\Lambda_{n]}} := \bigoplus_{\omega\in \Omega_{n]}}\left(D_{\omega_{x_0}}\bigotimes_{k=0}^{n-1}\bigotimes_{x\in\overrightarrow \Lambda_k}  D_{\omega_{\{x\}\cup \overrightarrow S(x)}}\right)
\end{equation}
Since the sequence $\{\mathcal D_{\Lambda_{n]}}\}_{n\in \mathbb N}$ is increasing and maximal abelian in $ \mathcal N_{\Lambda_{n]}}$ then the limit
$$
\mathcal D :=\overline{ \lim_{n\to \infty} \mathcal D_{\Lambda_{n]}}}^{C^\ast}
$$
is a diagonal algebra of $\mathcal N$.

By $h_{\mathcal N_{\Lambda_{n]}}}$ we denote a potential
associated to the restriction $\varphi\lceil{\mathcal
N_{\Lambda_{n]}}}$. Namely,
$$
\varphi_{\big \lceil \mathcal N_{\Lambda_{n]}}}(\cdot ) = \tr
\left(e^{-h_{\mathcal N_{\Lambda_{n]}}}\cdot}\right)
$$
One gets
$$
e^{-h_{\mathcal N_{\Lambda_{n]}}}} = E^{\mathcal{A}_{\Lambda_{n]}}}_{N_{\Lambda_{n]}}}(e^{-h_{\mathcal{A}_{\Lambda_{n]}}}})
$$

Then the potential $h_{\mathcal N_{\Lambda_{n]}}}$ has the following decomposition

\begin{equation}
h_{\mathcal N_{\Lambda_{n]}}} = K_{x_{0}} +
\sum_{k=1}^{n-1}\sum_{x\in \Lambda_n} H_{\{x\}_\cup
\overrightarrow S(x)}  + \widehat K_{n  }
\end{equation}
where
\begin{equation}
 K_{x_0} := - \sum_{\omega_{x_0}\in\Omega_{x_0}}\ln\bigl( \tr_{N^{x_0}_{\omega_{x_0}}}e^{-h^{x_0}_{\omega_{x_0}}}\bigr)P^{x_0}_{\omega_{x_0}}
\end{equation}
and for all $y\in \Lambda_n$,
$$
 \widehat K_{y}: = - \sum_{\omega_{y}\in\Omega_{y}}\ln\bigl( \tr_{\bar N^{y}_{\omega_{y}}}e^{-h^{y}_{\omega_{y}}}\bigr)P^{y}_{\omega_{y}},
$$
 and yet
$$
\widehat K_{\Lambda_n} := \bigotimes_{y\in \Lambda_n}\widehat K_{y} = -
\sum_{\omega_{\Lambda_n}\in \Omega_{\Lambda_n}}\bigotimes_{y\in \Lambda_n}\ln\bigl( \tr_{\bar N^{y}_{\omega_{y}}}e^{-h^{y}_{\omega_{y}}}\bigr)P^{y}_{\omega_{y}}.
$$

Let $\mathbb E_{\Lambda_n]}: \mathcal N_{\Lambda_n} \to \mathcal D_{\Lambda_n}$ be the canonical conditional expectation from $\mathcal N_{\Lambda_{n]}}$ into its maximal abelian subalgebra $\mathcal D_{\Lambda_{n]}}$.
Since  $\mathbb E_{\Lambda_{n]}}$ is trace-preserving then
\begin{equation}\label{phi_E_D}
\varphi_{\big\lceil \mathcal N_{\Lambda_{n]}}} =  \tr_{ \mathcal
N_{\Lambda_{n]}}}\left( e^{-h_{\mathcal N_{\Lambda_{n]}}}}\mathbb
E_{\Lambda_n]}(\cdot)\right)
\end{equation}

One the other hand, one has the  following compatibility conditions
\begin{equation}\label{compatibilityE}
{E_{\Lambda_{n+1]}}}_{\big\lceil{\mathcal{A}_{n]}}} = E_{\Lambda_{n]}} \quad ; \quad {\mathbb E_{\Lambda_{n+1]}}}_{\big\lceil{\mathcal N_{\Lambda_{n]}}}} = \mathbb E_{\Lambda_n]}
\end{equation}

This leads to the introduction of the following conditional expectation
\begin{equation}
  \mathfrak{E}_{\Lambda_{n]}} = \mathbb E_{\Lambda_n}\circ E_{\Lambda_{n]}}
\end{equation}
where $E_{\Lambda_{n]}}$ is given by \eqref{E_Vambad_n}. Using \eqref{compatibilityE},  it follows that
$${\mathfrak  E_{\Lambda_{n+1]}}}_{\big\lceil{\mathcal{A}_{n]}}} = \mathfrak  E_{\Lambda_{n]}}$$
Therefore, the limit
$$\mathfrak E : = \lim_{n\to \infty}\mathfrak  E_{\Lambda_{n]}}$$
exists in the strongly finite sense, and $\mathfrak E$ is a Umegaki conditional expectation  of $\mathcal{A} $ onto $\mathcal D$.

Let $\mu$ be the probability measure on $\mathrm{spec}(\mathcal D)$ associated to $\varphi_{\big\lceil{\mathcal D}} =: \varphi_\mu$.

The assertion (ii) of  Proposition \ref{prop_phiE} together with
\eqref{phi_E_D} implies
$${\varphi_{\mu}}_{\lceil \mathcal D_{\Lambda_{n]}}} \circ \mathfrak{E}_{\Lambda_{n]}} = {\varphi_{\mu}}_{\lceil \mathcal D_{\Lambda_{n]}}} $$
Therefore, a standard continuity argument yields to $\varphi_\mu\circ\mathfrak E = \varphi_\mu$.
The markovianity of the measure  $\mu$  will be proved in the next section.

\section{A classical Markov chain on the spectrum of the diagonal algebra}

This section is devoted to the proof of the last part of Theorem
\ref{main_thm} which concerns the Markov nature of the classical
probability measure $\mu$ canonically associated with the
restriction $\varphi_\mu := \varphi\lceil_{\mathcal D}$ of the QMS
$\varphi$ considered in Section \ref{diag_section}.

The diagonal sub-C$^\ast$--algebra $\mathcal D$ of $\mathcal A$ is
isomorphic to the algebra of all bounded complex valued functions
$C(K)$, where $K$ is a compact Hausdorff space of all maximal
ideals of $\mathcal D$.

Taking into  account (\ref{decomp_D_Vn}), the spectrum
$\mathrm{spec}(\mathcal D_{\Lambda_{n]}})$ is given by  the
following disjoint union
$$
\mathrm{spec}(D_{\Lambda_{n]}}) = \bigcup_{\omega_{\Lambda_{n]}}\in \Omega_{\Lambda_{n]}}} \mathcal S_{\omega_{x_0}} \times \big( \prod_{k=0}^{n-1}\prod_{x\in \Lambda_k}\mathcal S_{\omega_{\{x\}\cup \overrightarrow S(x)}}\big)
$$
where $\mathcal S_{\omega_{x_0}}$ and $\mathcal S_{\omega_{\{x\}\cup \overrightarrow S(x)}}$ denote $\mathrm{spec}(D_{\omega_{\{x_0\}}})$
 and $\mathrm{spec}(D_{\omega_{\{x\}\cup \overrightarrow S(x)}})$ respectively.

Let $A\in \mathrm{spec}(\mathcal D)$ then
\begin{equation}\label{integ_decomp}
\varphi(A) = \int \mathfrak E(A)(\omega) \mu(d\omega)
\end{equation}

For a given event $A\in \mathrm{spec}({D_{\Lambda_{n}}})$ the past
is $\mathrm{spec}({D_{\Lambda_{n-1]}}})$ and the future is
$\mathrm{spec}(D_{[n+1})$. Then the Markov property \cite{[Spi]}
reads
\begin{equation}\label{classical_MP}
  \mathbb{P}(\omega_{\Lambda_{n+1}} \mid \bar\omega_{\Lambda_{n }}, \bar\omega_{\Lambda_{n-1}}, \dots,  \bar\omega_{\Lambda_0} ) = \mathbb{P}(\omega_{\Lambda_{n+1}} \mid \bar\omega_{\Lambda_n}), \quad \forall \omega_{\Lambda_{n+1}}\in\Omega_{n+1}
\end{equation}
where $\bar\omega_{\Lambda_j}\in \Omega_j, \,  j=1,\cdots,n$
satisfying
 $$ \mathbb{P}( \bar\omega_{\Lambda_{n}}, \bar\omega_{\Lambda_{n-1}}, \dots, \bar\omega_{\Lambda_0} ) >0.$$

Consider
$$
f =  \sum_{(\Lambda_x)_{x\in \Lambda_{n]}}} \chi_{\mathcal S_{\omega_{x_0}}\times \prod_{k=0}^{n-1}\prod_{x\in \Lambda_k}\mathcal S_{\omega_x, \omega_{\overrightarrow{S}(x)}}} f_{\omega_{x_0}}\otimes\bigotimes_{k=0}^{n-1}\bigotimes_{x\in \Lambda_n}f_{\omega_x, \omega_{\overrightarrow{S}(x)}} \in \mathcal D_{\Lambda_{n]}}
$$
One has
\begin{equation}\label{phu(f)}
  \varphi(f) =  \sum_{(\Lambda_x)_{\{x\in \Lambda_{n]}\}}} \big(\int_{\mathcal S_{\omega_{x_0} }}T^{(x_0)}_{\omega_{x_0} } \big)\times \prod_{k=0}^{n-1}\prod_{x\in \Lambda_k}\big(\int_{\mathcal S_{\omega_x, \omega_{\overrightarrow{S}(x)} }}T^{(x)}_{\omega_x, \omega_{\overrightarrow{S}(x)} } \big)
\end{equation}
where $T^{(x_0)}_{\omega_{x_0} }$ and $T^{(x)}_{\omega_x,
\omega_{\overrightarrow{S}(x)} }$ are positive densities, and
$\int$ assigns weight $1$ to the minimal projections
$P^{(x_0)}_{\omega_{x_0}}$ and
$P_{\omega_{x\cup\overrightarrow{S}(x)}}^{(x)}$.

Let $\bar\omega_0 \in\mathrm{spec}(D_{\Lambda_n}),  \cdots, \bar\omega_n \in\mathrm{spec}(D_{\Lambda_n})$
$$
P^{(x)}_{\bar\omega_x} = \sum_{\omega_{n-1]}} \bigotimes_{k=0}^{n-2}\bigotimes_{x\in \Lambda_k}\chi_{\mathcal S^{(x)}_{\omega_x, \omega_{\overrightarrow{S}(x)}}} \otimes \bigotimes_{y\in \Lambda_{n-1}}\chi_{\mathcal S^{(y)}_{\omega_y, \bar\omega_{\overrightarrow{S}(y)}}}
$$
Inside $D_{\Lambda_{n]}}$, one has
$$
\chi_AP_{\bar\omega_n}^{(n)} = \sum_{a\in A}\bigotimes_{y\in \Lambda_{n}}\chi_{\{ a_{\omega_y(a) , \omega_{\overrightarrow{S}(y)}(a)}\}}
$$
where
$$a = \prod_{y\in \Lambda_n}a_{\omega_y(a) , \omega_{\overrightarrow{S}(y)}(a)}.$$
We have the following computations

\begin{eqnarray*}
\mathbb P(\bar\omega_{\Lambda_n})& = & \varphi\left(P_{\bar\omega_n}^{(n)}\right)\\
& =& \sum_{\omega_{\Lambda_0}, \cdots, \omega_{\Lambda_{n-1}}} \big(\int_{\mathcal S_{\omega_{x_0} }}T^{(x_0)}_{\omega_{x_0} } \big)\left(\prod_{k=0}^{n-2}\prod_{x\in \Lambda_k} \int T^{(x)}_{\mathcal S^{(x)}_{\omega_x, \omega_{\overrightarrow{S}(x)}}}\right)\left( \prod_{y\in \Lambda_{n-1}}\int T^{(y)}_{\mathcal S^{(y)}_{\omega_y, \bar\omega_{\overrightarrow{S}(y)}}}\right)
\end{eqnarray*}
\begin{eqnarray*}
\mathbb P(\bar\omega_{\Lambda_n},  \omega_{\Lambda_{n+1}})& = & \varphi\left(P_{\bar\omega_n}^{(n)}\otimes P_{\omega_{n+1}}^{(n+1)}\right)\\
& =& \left[\sum_{\omega_{\Lambda_0}, \cdots, \omega_{\Lambda_{n-1}}} \left(\int_{\mathcal S_{\omega_{x_0} }}T^{(x_0)}_{\omega_{x_0} } \right)\left(\prod_{k=0}^{n-2}\prod_{x\in \Lambda_k} \int T^{(x)}_{\mathcal S^{(x)}_{\omega_x, \omega_{\overrightarrow{S}(x)}}}\right)\left( \prod_{y\in \Lambda_{n-1}}\int T^{(y)}_{\mathcal S^{(y)}_{\omega_y, \bar\omega_{\overrightarrow{S}(y)}}}\right)\right]\\
&&\times \left( \prod_{z\in \Lambda_n}\int T^{(z)}_{\mathcal S^{(z)}_{\bar\omega_z,  \omega_{\overrightarrow{S}(z)}}}\right).
\end{eqnarray*}

On the other hand one has
\begin{eqnarray*}
\mathbb P( \bar\omega_{\Lambda_0}, \cdots, \bar\omega_{\Lambda_n}) &=& \varphi\left( P^{(0)}_{ \bar\omega_{\Lambda_0}}\otimes        \cdots P^{(n)}_{ \bar\omega_{\Lambda_n}}\right) \\
     &=&  \left(\int_{\mathcal S_{\bar\omega_{x_0} }}T^{(x_0)}_{\bar\omega_{x_0} } \right)\left(\prod_{k=0}^{n-1}\prod_{x\in \Lambda_k} \int T^{(x)}_{\mathcal S^{(x)}_{\bar\omega_x, \bar\omega_{\overrightarrow{S}(x)}}}\right). \\
\end{eqnarray*}

\begin{eqnarray*}
\mathbb P( \bar\omega_{\Lambda_0}, \cdots, \bar\omega_{\Lambda_n}, \omega_{\Lambda_{n+1}}) &=& \varphi\left( P^{(0)}_{ \bar\omega_{\Lambda_0}}\otimes        \cdots P^{(n)}_{ \bar\omega_{\Lambda_n}}\right) \\
     &=&  \left[ \left(\int_{\mathcal S_{\bar\omega_{x_0} }}T^{(x_0)}_{\bar\omega_{x_0} } \right)\left(\prod_{k=0}^{n-1}\prod_{x\in \Lambda_k} \int T^{(x)}_{\mathcal S^{(x)}_{\bar\omega_x, \bar\omega_{\overrightarrow{S}(x)}}}\right)\right]\times \left( \prod_{z\in \Lambda_n}\int T^{(z)}_{\mathcal S^{(z)}_{\bar\omega_z,  \omega_{\overrightarrow{S}(z)}}}\right). \\
\end{eqnarray*}
Recapitulating, one gets
$$
\frac{\mathbb P(\bar\omega_{\Lambda_n},
\omega_{\Lambda_{n+1}})}{\mathbb P(\bar\omega_{\Lambda_n})} =
\prod_{z\in \Lambda_n}\int T^{(z)}_{\mathcal
S^{(z)}_{\bar\omega_z,  \omega_{\overrightarrow{S}(z)}}}  =
\frac{\mathbb
P(\bar\omega_{\Lambda_0},\cdots,\bar\omega_{\Lambda_n},
\omega_{\Lambda_{n+1}})}{\mathbb
P(\bar\omega_{\Lambda_0},\cdots,\bar\omega_{\Lambda_n})}.
$$
This leads to (\ref{classical_MP}). This completes the proof.




\begin{thebibliography}{20}
\bibitem{[Ac74f]}
 Accardi L., Noncommutative Markov chains, \emph{Proc. of Int. School of Math. Phys.
 Camerino } (1974), 268--295.


\bibitem{ACe} Accardi L., Cecchini C., Conditional expectations in von Neumann algebras and a Theorem of Takesaki, \textit{J. Funct. Anal.}
{\bf 45} (1982), 245--273

\bibitem{AccFi}  Accardi L., Fidaleo F., Non homogeneous quantum Markov states and quantum Markov fields, \textit{J. Funct. Anal.} 200 (2003), 324-347.

\bibitem{AccFid03} Accardi L., Fidaleo F., Quantum Markov fields, \textit{Inf. Dim.
Analysis, Quantum Probab. Related Topics} {\bf  6} (2003),
123-138.


\bibitem{[AcFiMu07]}
Accardi L., Fidaleo F. Mukhamedov, F., Markov states and chains on
the CAR algebra, \textit{Inf. Dim. Analysis, Quantum Probab. Related
Topics} {\bf 10} (2007), 165--183.


\bibitem{[AcFr80]}
Accardi L., Frigerio A., Markovian cocycles, \textit{Proc. Royal
Irish Acad.} {\bf 83A} (1983) 251-263.

\bibitem{AccKhreOhy} Accardi L.,  Khrennikov A., Ohya M., Quantum Markov Model for Data from Shar-Tversky Experiments in
Cognitive Psychology, \textit{Open Systems \& Information
Dynamics} {\bf 16}(2009), 371-385.


\bibitem{AccMuSa1} Accardi L., Mukhamedov, F. Saburov M. On Quantum Markov Chains on Cayley tree I:
 uniqueness of the associated chain with XY -model on the Cayley tree of order two,
 \textit{Inf. Dim. Analysis, Quantum Probab. Related Topics } 14(2011), 443--463.


\bibitem{AccMuSa2} Accardi L., Mukhamedov, F. Saburov M. On Quantum Markov Chains on Cayley tree II:
 Phase transitions for the associated chain with XY -model on the Cayley tree of order three,
  \textit{Ann. Henri Poincare } 12(2011), 1109-1144.

\bibitem{AccMuSa3} Accardi L., Mukhamedov, F. Saburov M. On Quantum Markov Chains on Cayley tree III:
 Ising model, \textit{Jour. Statis. Phys.} 157 (2014), 303-329.




\bibitem{AOM} Accardi L., Ohno, H., Mukhamedov F., Quantum Markov fields on
graphs, \textit{Inf. Dim. Analysis, Quantum Probab. Related Topics}
{\bf 13}(2010), 165--189.


  \bibitem{[AccMuSou]}
Accardi L., Mukhamedov F., Souissi A., Construction of a new class
of quantum Markov fields, \textit{Adv. Oper. Theory} 1 (2016), no.
2, 206-218.


\bibitem{BR} Bratteli O., Robinson D.W., \textit{Operator algebras and quantum statistical
mechanics I}, Springer-Verlag, New York, 1987.

\bibitem{CDS} Chakrabarti B.K., Dutta A., Sen P., \textit{Quantum Ising phases and
transitions in transverse Ising models}, Springer, Berlin, 1996

\bibitem{CV}  Cirac J.I., Verstraete F. Renormalization and tensor product states in spin chains and
lattices, \textit{J. Phys. A. Math. Theor.} {\bf 42} (2009),
504004.

\bibitem{D2010} Dorogovtsev S.N., \textit{Lectures on Complex Networks},  (Oxford Master
Series in Statistical, Computational, and Theoretical Physics),
Oxford Univ. Press 2010.



\bibitem{fannes}
Fannes M., Nachtergaele B. Werner R. F., Ground states of VBS
models on Cayley trees, \textit{J. Stat. Phys.} {\bf 66} (1992)
939--973.

\bibitem{fannes2}
Fannes M., Nachtergaele B. Werner R. F., Finitely correlated
states on quantum spin chains, \textit{Commun. Math. Phys.} {\bf
144} (1992) 443--490.


\bibitem{FF_FM}   Fidaleo F.,  Mukhamedov, F. Diagonalizability of non homogenuous quantum Markov states  and associated von Neumann
algebras, \textit{Probab. Math. Stat.} {\bf 24} (2004), 401-418.


\bibitem{GZ} Golodets V.Y., Zholtkevich G.N. Markovian KMS states, \textit{Theor. Math.
Phys.} {\bf 56}(1983), 686-690.


\bibitem{Kum} K\"{u}mmerer B. Quantum Markov processes and applications in physics. In book: Quantum independent increment processes. II,  259--330, \textit{ Lecture Notes in Math.}, 1866, Springer, Berlin, 2006.

\bibitem{L86} Liebmann R. \textit{Statistical mechanics of periodic frustrated Ising
systems}, Springer, Berlin, 1986

\bibitem{Lib01} Liebscher V. Markovianity of quantum random fields, \textit{Proceedings Burg Conference
15-20 March 2001, W. Freudenberg (ed.)}, World Scientific, QP-PQ
Series 15 (2003) 151-159.



\bibitem{MS} Moessner R., Sondhi S.L., Ising models of quantum frustrations,
\textit{Phys. Rev. B} {\bf 63}(2001), 224401.


\bibitem{MBS161} Mukhamedov F., Barhoumi A., Souissi A., Phase transitions for quantum Markov chains associated with Ising type models on a Cayley tree, \textit{J. Stat. Phys.} {\bf 163} (2016), 544--567.

\bibitem{MBS162} Mukhamedov F., Barhoumi A., Souissi A., On an algebraic property of the disordered phase of
the Ising model with competing interactions on a Cayley tree,
\textit{Math. Phys. Anal. Geom.} {\bf 19}(2016), 21.

\bibitem{MGS17} Mukhamedov F., El Gheteb S., Uniqueness of quantum Markov chain associated with
XY-Ising model on the Cayley tree of order two, \textit{Open Sys.
\& Infor. Dyn.} {\bf 24} (2017), no. 2, 175010.

\bibitem{MGS19} Mukhamedov F., El Gheteb S., Clustering property of Quantum Markov Chain associated
to XY-model with competing Ising interactions on the Cayley tree
of order two, \textit{Math. Phys. Anal. Geom.} {\bf 22}(2019), 10.


\bibitem{[MuSou18]}
Mukhamedov F., Souissi A.,  Quantum Markov States on Cayley trees,
\textit{J. Math. Anal. Appl.} {\bf 473}(2019), 313-333.






\bibitem{Or} Orus R. A practical introduction of tensor networks: matrix
product states and projected entangled pair states, \textit{Ann of
Physics} {\bf 349} (2014) 117-158.


\bibitem{[Spi]}
Spitzer F., Markov random fields on an infinite tree, \textit{Ann.
Prob.} {\bf 3} (1975) 387-398.

\bibitem{Str} Stratilla S., Modular theory in operator algebras, \textit{Abacus Press, Tunbridge Wells,
Kent,} 1981.



\end{thebibliography}
\end{document}